# Optimal control for manipulating vibrational wave packets through polarizability interactions induced by non-resonant laser pulses

Reon Ishii[1], Tomotaro Namba[1], Hiroyuki Katsuki[2], Kenji Ohmori[3, 4], and Yukiyoshi Ohtsuki*[1]

1) Department of Chemistry, Graduate School of Science, Tohoku University
6-3 Aramaki Aza-Aoba, Aoba-ku, Sendai 980-8578, Japan

2) Graduate School of Science and Technology, Nara Institute of Science and Technology
8916-5 Takayama-cho, Ikoma, Nara 630-0192, Japan

3) Institute for Molecular Science, National Institutes of Natural Sciences
Myodaiji, Okazaki 444-8585, Japan

4) The Graduate University for Advanced Studies (SOKENDAI)
Shonan Village, Hayama, Kanagawa 240-0193, Japan

* Corresponding author: yukiyoshi.ohtsuki.d2@tohoku.ac.jp

**Abstract**

On the basis of optimal control theory, we numerically study how to optimally manipulate molecular vibrational dynamics by using cycle-averaged polarizability interactions induced by mildly intense non-resonant laser (NR) pulses. As the essential elements to be controlled are the probability amplitudes, namely, the populations and the relative phases of the vibrational eigenstates, we consider three fundamental control objectives: selective population transfer, wave packet shaping that requires both population control and relative-phase control, and wave packet deformation suppression that solely requires relative-phase control while avoiding population redistribution. The non-trivial control of wave packet deformation suppression is an extension of our previous study on wave packet spreading suppression. Focusing on the vibrational dynamics in the $B$ state of $I_2$ as a case study, we adopt optimal control simulations and model analyses under the impulsive excitation approximation to systematically examine how to achieve the control objectives with shaped NR pulses. Optimal solutions are always given by NR pulse trains, in which each pulse interval and each pulse intensity are adjusted to cooperate with the vibrational dynamics to effectively utilize the quantum interferences to realize the control objectives with high probability.

**I. Introduction**

The coherent control of vibrational wave packets of molecules through electronic and/or vibrational excitations has attracted much attention as it directly manipulates wave functions to achieve, for example, specified nuclear configurations associated with target reaction dynamics [1-3]. Although we often adopt resonant laser pulses for this purpose, optical selection rules impose restrictions on optically accessible regions so that in the optically "dark" regions, the vibrational wave packets evolve in time in the absence of control pulses. This puts a limit on the capability of the resonant laser pulses to control the vibrational wave packets because physically and/or chemically important processes such as nonadiabatic transitions often occur in the optically "dark" regions [4]. Mildly intense non-resonant laser (NR) pulses are considered an alternative control tool as they introduce induced dipole interactions, that is, polarizability interactions, which are almost unrestricted by the optical selection rules. The trade-off in NR pulse control is that the polarizability interactions are cycle-averaged over the optical frequencies, resulting in the loss of optical-frequency-dependent control knobs. To overcome the lack of control knobs, the NR pulse envelopes should be suitably shaped to effectively cooperate with the vibrational wave packets to achieve the control objectives.

The cycle-averaged polarizability interactions manipulate the vibrational wave packets through the dynamic Stark shifts (energy shifts) and the Raman transitions even though they are inseparable from each other [5-16]. Examples include the so-called dynamic Stark control of reaction dynamics [5-12], which mainly utilizes the dynamic Stark shifts to distort the potential energy surfaces to selectively enhance/suppress the specified reactions. The population transfer achieved by the stimulated Raman transitions has been studied particularly in the adiabatic excitation regimes because of the robustness [17-22]. Typical examples with and without the (near) resonant intermediate states include the stimulated Raman adiabatic passage (STIRAP) [17, 18] and the Stark-induced adiabatic Raman passage (SARP) [19-21], which are often discussed by using the three-state lambda and the two-level systems, respectively. Another example is the amplification of torsional motion by impulsive Raman transitions, the effectiveness of which has been demonstrated in a variety of systems [13-16]. Recently, we proposed a control scheme for suppressing vibrational wave packet spreading (dephasing) [23, 24] by periodic NR pulse irradiation, although we restricted ourselves to the vibrational wave packets created by the Fourier transform-limited (TL) pump pulses, which are characterized by zero initial relative phases.

This control is non-trivial because the relative phases among the vibrational eigenstates are solely adjusted while avoiding the population redistribution during the control period so that the controlled wave packets evolve in time as if they are in a harmonic potential. Note that the intensity of the NR pulses used in the dephasing suppression is not sufficient to distort the potential energy surface. In addition to the vibrational dynamics, the (shaped) NR pulses are also applied to molecular rotational dynamics in which the rotational wave packets are controlled mainly by the rotational Raman transitions [25-35]. Examples include molecular alignment control and the realization of specified rotational population distributions [34, 35].

Motivated by recent developments in the coherent control with NR pulses, in this study, we systematically examine how effectively the NR pulses control the probability amplitudes in non-adiabatic regimes on the basis of quantum optimal control theory [36-38] through a case study of vibrational dynamics. Because of experimental feasibility, we consider the vibrational dynamics in the *B* state of $I_2$ as in the previous study. We discuss the effectiveness and the limitations of NR pulse control by focusing on three fundamental objectives: (i) non-adiabatic selective population transfer, (ii) wave packet shaping that requires both population and relative-phase control, and (iii) wave packet deformation suppression that solely requires relative-phase control without changing the population distribution. The non-trivial control of wave packet deformation suppression is an extension of our previous study on wave packet spreading suppression. Specifically, we require the wave packet, which moves in an anharmonic potential, to periodically oscillate as if it is in a harmonic potential and therefore, the wave packet recovers its original shape in (almost) every vibrational period. The control mechanisms are discussed on the basis of the structures of the numerically designed optimal NR pulses and the model analyses under the impulsive excitation approximation [39, 40].

In Sec. II, we outline the optimal control simulation with the polarizability interactions, the numerical details of which are provided in Appendices A and B. In Sec. III, we show the numerical results associated with the selective population transfer, the wave packet shaping, and the suppression of wave packet deformation. We also adopt an impulsive excitation model to qualitatively discuss the control mechanisms. We summarize the present study in Sec. IV.

**II. Optimal control simulation with NR pulses**

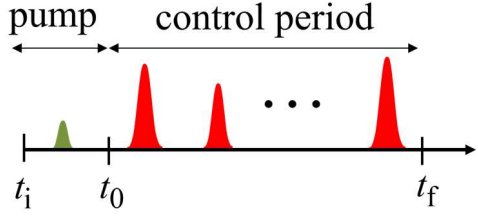

Figure 1
Schematic illustration of laser pulse sequence during the pump period $[t_i, t_0]$ and the control period $[t_0, t_f]$.

      We consider a two-electronic-state system composed of the $X$ and $B$ electronic states of $I_2$, which is assumed to be initially in the lowest vibrational state $|0_X\rangle$ in the ground $X$ electronic state. As illustrated in Fig. 1, we first prepare specified initial vibrational states in the $B$ electronic state by adjusting the shapes of weak pump pulses $E_{\text{pump}}(t)$ during the pump period $[t_i, t_0]$. We briefly summarize the initially excited states and the molecular parameters in Appendix A. By suitably choosing the optical frequency, we can safely assume that the NR pulses do not induce electronic transitions so that the initially excited states evolve in time within the $B$ state potential while being controlled by the linearly polarized NR pulses during the control period $[t_0, t_f]$. Because of this, the time evolution of the vibrational wave packet in the $B$ state is described by the Schrödinger equation

$$i\hbar \frac{\partial}{\partial t}|\psi(t)\rangle = H_B(t)|\psi(t)\rangle = \left\{ H_B^0 - \frac{1}{4}\alpha_B(r)[\mathcal{E}_{\text{NR}}(t)]^2 \right\}|\psi(t)\rangle, \qquad (1)$$

where the (normalized) initial state in the $B$ state is given by $|\psi(t_0)\rangle$ (Appendix A). Here, the $B$ state vibrational Hamiltonian $H_B(t)$ is composed of the field-free part $H_B^0$ and the polarizability interaction cycle-averaged over the optical frequency of the NR pulse. In Eq. (1), $\mathcal{E}_{\text{NR}}(t)$ and $\alpha_B(r)$ are the envelope of the NR pulse and the polarizability function with $r$ being the internuclear distance, respectively. The vibrational eigenvalue $\hbar\omega_v$ and eigenstate $|v\rangle$ are defined by $H_B^0|v\rangle = \hbar\omega_v|v\rangle$, where $v$ is the vibrational quantum number.

      We consider the general optimal control problem [36, 41] defined by

$$J = \langle \psi(t_f) | X | \psi(t_f) \rangle + \int_{t_0}^{t_f} dt \langle \psi(t) | Y(t) | \psi(t) \rangle, \quad (2)$$

which is expressed in terms of the target operators $X$ and $Y(t)$ that quantify our physical objective at the final time $t_f$ and during the control period $[t_0, t_f]$, respectively. If we apply the calculus of variations to Eq. (2) subject to the constraint given by Eq. (1), we will obtain the maximal condition

$$\mathrm{Im} \langle \xi(t) | \alpha_B(r) | \psi(t) \rangle \mathcal{E}_{\mathrm{NR}}(t) = 0. \quad (3)$$

Here, the Lagrange multiplier $|\xi(t)\rangle$ associated with the constraint [Eq. (1)] obeys the equation of motion

$$i\hbar \frac{\partial}{\partial t} |\xi(t)\rangle = H_B^\dagger(t) |\xi(t)\rangle - i\hbar Y(t) |\psi(t)\rangle, \quad (4)$$

with the final condition $|\xi(t_f)\rangle = X |\psi(t_f)\rangle$.

The so-called coupled pulse-design equations composed of Eqs. (1), (3), and (4) are solved iteratively by using the previously developed monotonically convergent algorithm that can deal with the nonlinear interaction with respect to $\mathcal{E}_{\mathrm{NR}}(t)$ [42-44]. Note that we introduce the instantaneous penalty [45] through the polarizability interaction to suppress the intensities of the optimal NR pulses so that we replace $\alpha_B(r)$ with $\alpha_\gamma(r) = (1 + i\gamma)\alpha_B(r)$ when solving the coupled pulse-design equations. We choose the value of the positive constant parameter $\gamma$ to adjust the pulse intensity. After obtaining the optimal NR pulses in the presence of the instantaneous penalty, we substitute the optimal NR pulses into the original equation of motion [Eq. (1)] to calculate the dynamics without being affected by the artificial parameter $\gamma$.

## III. Results and discussion

Three kinds of control objectives are considered in the present study to illustrate how the NR pulses control the probability amplitudes in the vibrational dynamics. The first two control objectives are (i) the selective population transfer (Sec. IIIA) that solely requires the population control, and (ii) the wave packet shaping (Sec. IIIC) that requires the control of both the population distribution and the relative phases. The third control objective is (iii) the suppression of the wave packet deformation (Sec. IIID), in which we need to solely control the relative phases without changing the population distribution. In the first two applications, the system is assumed to be initially prepared in a specified vibrational eigenstate $|\psi(t_0=0)\rangle = |v=30\rangle$, whereas in the third application, various pump pulses are used to prepare the specified initial vibrational wave packets.

In the following numerical examples, we will use the dimensionless envelope function $f(t)$ defined by $\mathcal{E}_{\text{NR}}(t) = \mathcal{E}_{\text{NR}}^0 f(t)$. Referring to the previous study [46], we measure the magnitude of the interaction in units of $V_\alpha(r_e) = \alpha_B(r_e)(\mathcal{E}_{\text{NR}}^0)^2 = 1.3 \times 10^{-3}$ a.u. with $r_e$ being the equilibrium internuclear distance in the $B$ state. Here, the polarizability interaction expressed in terms of $f(t)$ is given by $-\alpha_B(r)[\mathcal{E}_{\text{NR}}(t)]^2/4 = -V_\alpha(r)[f(t)]^2/4$.

**A. Selective population transfer**

We start the discussion with the selective population transfer, in which the temporal width of the pump pulse is assumed to be sufficiently long to initially prepare the specified vibrational eigenstate $|\psi(t_0=0)\rangle = |v=30\rangle$. Our aim is to achieve the selective population transfer into the state $|v=31\rangle$ within the specified control period $[t_0=0, t_f]$. The target operators in Eq. (2) are set to $X = |v=31\rangle\langle v=31|$ and $Y(t) = 0$. One of the difficulties of the selective population transfer originates in the frequency resolution associated with the adjacent vibrational Raman transitions that are close to each other in energy. For example, the present system gives $\omega_{31,30} = 69.1\,\text{cm}^{-1}$ and $\omega_{30,29} = 71.3\,\text{cm}^{-1}$, where $\omega_{v+1,v} \equiv \omega_{v+1} - \omega_v$. Distinguishing the two Raman transition frequencies $\omega_{31,30}$ and $\omega_{30,29}$ takes ca. $33 T_{31,30} \simeq 15.8\,\text{ps}$ with $T_{31,30} = 485\,\text{fs}$, where the vibrational period is defined by $T_{v+1,v} = 2\pi/\omega_{v+1,v}$.

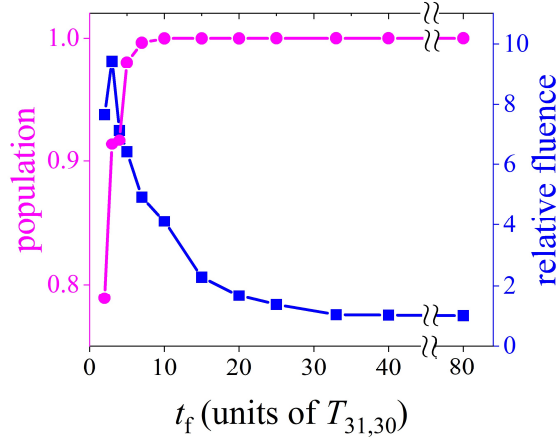

Figure 2
Population of the target state $|v=31\rangle$ as a function of the final time $t_\mathrm{f}$ (purple line) measured in units of the vibrational period defined by $T_{31,30} = 482$ fs. The blue line shows the relative pulse fluence with respect to the fluence in the case of $t_\mathrm{f} = 80 T_{31,30}$, i.e., the Raman π pulse.

With this in mind, we numerically calculate the optimal NR pulse envelopes that maximally achieve the selective population transfer for several final times $t_\mathrm{f}$. We see from Fig. 2 that the optimal NR pulses achieve almost complete population transfer even when $t_\mathrm{f} \sim 10 T_{31,30}$, which is much shorter than that expected from the frequency resolution, i.e., $t_\mathrm{f} \geq 33 T_{31,30}$. Figure 2 also shows the relative pulse fluence with respect to the fluence associated with the so-called Raman π pulse, where the Raman pulse area is defined by the temporal integral of the polarizability interaction over the control period $[t_0, t_\mathrm{f}]$ divided by $\hbar$ [47]. The relative fluence rapidly decreases as $t_\mathrm{f}$ increases and becomes almost constant (the Raman π pulse) after $t_\mathrm{f} = 33 T_{31,30}$.

Before examining the non-trivial control mechanisms in the case of $t_\mathrm{f} < 33 T_{31,30}$, we consider the results when $t_\mathrm{f} = 33 T_{31,30}$ because the control mechanisms are expected to be easier to understand than when $t_\mathrm{f} < 33 T_{31,30}$. Figure 3 shows (a) the envelope $f(t)$ of the optimal NR pulse and (b) the major populations as a function of time when $t_\mathrm{f} = 33 T_{31,30}$. We see from Fig. 3(a) that the optimal solution is the pulse train composed of 33 pulses appearing regularly at every vibrational period $\sim T_{31,30}$, which is indicated by the purple solid squares. As expected from the Fourier spectrum (not shown here) of the optimal pulse train, the initial population of the state $|v=30\rangle$ is gradually and selectively transferred to the target state $|v=31\rangle$ while only a small portion of the population is distributed in the other states during the control period [see

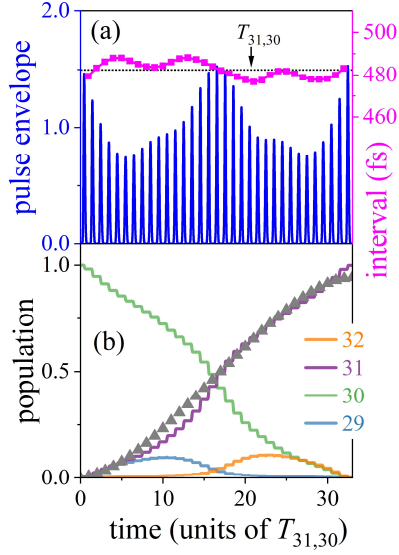

Figure 3
(a) Optimal pulse envelope as a function of time (blue line) when $t_\text{f} = 33 T_{31,30}$. The intervals between the adjacent pulses are shown by purple solid squares. The black dashed line shows $T_{31,30} = 482$ fs. (b) Populations of four major vibrational eigenstates involved in the population transfer control. Gray solid triangles show the population of the target state $|v=31\rangle$, which is derived from the three-state model under the impulsive excitation approximation (see Sec. IIIB).

Fig. 3(b)]. The gray solid triangles show the population of the state $|v=31\rangle$ derived from the analytical model, which will be discussed in the next subsection. If we scrutinize the structure of the pulse train, we see that pulses with slightly high intensities appear around both ends and the middle of the control period. The wave packet around the initial (final) time is approximated by the vibrational eigenstate $|v=30\rangle$ ($|v=31\rangle$), which minimizes the interference effects between $|v=30\rangle$ and $|v=31\rangle$. The slightly intense pulses are required to overcome the minimal quantum interference effects to induce the population transfer. Contrary to this, the intense pulses around the middle of the control period actively utilize the interference effects to efficiently induce the population transfer.

We next consider $t_\text{f} = 25 T_{31,30}$ as an example of the non-trivial population transfer control, the results of which are summarized in Fig. 4. Although the optimal pulse in Fig. 4(a) looks like a simple pulse train composed of 25 pulses, the intervals between the adjacent pulses are no longer constant as indicated by the solid purple squares. Roughly speaking, the intervals longer than $\sim T_{31,30}$ appear in the first half and those close to $\sim T_{30,29}$ (=468 fs) appear

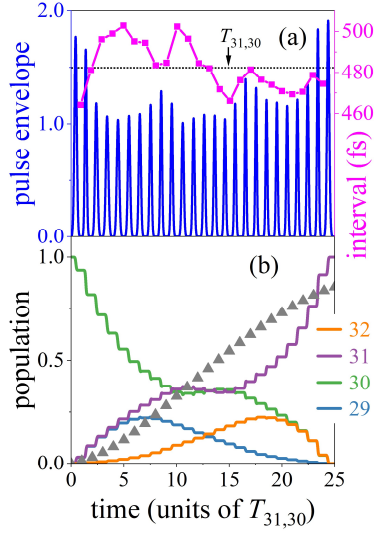

Figure 4
(a) Optimal pulse envelope as a function of time (blue line) when $t_f = 25 T_{31,30}$. The intervals between the adjacent pulses are shown by purple solid squares. The black dashed line shows $T_{31,30} = 482$ fs. (b) Populations of four major vibrational eigenstates involved in the population transfer control. Gray solid triangles show the population of the target state $|v = 31\rangle$, which is derived from the three-state model under the impulsive excitation approximation (see Sec. IIIB).

in the second half. The pulse train with irregular pulse intervals could lead to a broader Fourier spectrum than the pulse train with regular pulse intervals and could cause population transfer among the unwanted vibrational eigenstates. In fact, certain amounts of populations appear in states $|v = 29\rangle$ and $|v = 32\rangle$ during the control period in Fig. 4(b), although an almost perfect population transfer is achieved at the final time. We will examine these non-trivial control mechanisms by adopting the analytical model in the next subsection.

Finally, we note that when the control times are set to shorter than $t_f < 15 T_{31,30}$, highly structured optimal pulse trains (not shown here) with large pulse fluence are obtained, as shown in Fig. 2. The highly structured pulse trains would actively manipulate the quantum interferences through multiple Raman transitions to achieve the selective population transfer (Fig. 2) although it is difficult to interpret the control mechanisms.

**B. Three-state model with impulsive excitation approximation**

Referring to the numerically designed optimal NR pulses in Figs. 3(a) and 4(a), we assume that the pulse train is expressed as temporally separated $N$ pulses such that $\mathcal{E}_{\text{NR}}(t) = \sum_{n=1}^{N} \varepsilon_n(t - \tau_n)$ with the time delay $\{\tau_n\}$. If we further assume the impulsive excitation approximation, the wave packet at time $t_f$ is expressed as

$$|\psi(t_f)\rangle = U^{(0)}(t_f - \tau_N) e^{iR_N} U^{(0)}(\tau_N - \tau_{N-1}) \cdots e^{iR_n} U^{(0)}(\tau_n - \tau_{n-1}) \cdots e^{iR_1} U^{(0)}(\tau_1 - \tau_0) |\psi_0\rangle, \quad (5)$$

with $\tau_0 \equiv t_0$ and the initially excited state $|\psi(t_0)\rangle \equiv |\psi_0\rangle$. The free propagator and the operator $R$ associated with the polarizability interaction are, respectively, defined by

$$U^{(0)}(\tau_n - \tau_{n-1}) = \left( \frac{-iH_B^0(\tau_n - \tau_{n-1})}{\hbar} \right) \quad (n = 1, 2, \cdots, N) \quad (6)$$

and

$$R_n = \frac{1}{4\hbar} \int_{-\infty}^{\infty} dt \, U^{(0)\dagger}(t) \alpha_B(r) U^{(0)}(t) [\varepsilon_n(t)]^2. \quad (7)$$

As a minimal model of the present vibrational system, we consider a three-state model composed of the vibrational eigenstates $|v+1\rangle$, $|v\rangle$, and $|v-1\rangle$. In the matrix representation, we denote the initially excited state $|\psi_0\rangle$ and the wave packet at time $\tau_n$ (immediately after the $n$th pulse) $|\psi(\tau_n)\rangle$ as $\boldsymbol{\psi}_0$ and $\boldsymbol{\psi}(\tau_n) = [C_{v+1}(\tau_n) \; C_v(\tau_n) \; C_{v-1}(\tau_n)]^t$, respectively, where the superscript "$t$" means the transposed matrix. The operators $R_n$ and $U^{(0)}(\tau_n - \tau_{n-1})$ are expressed as $\boldsymbol{R}_n$ and $\boldsymbol{U}^{(0)}(\tau_n - \tau_{n-1})$, respectively. Because we can safely ignore the vibrational quantum number dependence of the dynamic Stark shifts in the vibrational eigenstates considered here, we will remove the shifts as the global phase. Then, we have

$$e^{i\mathbf{R}_n}\mathbf{U}^{(0)}(\tau_n - \tau_{n-1}) = \begin{bmatrix} A_n^{(+)}e^{-i\Phi_{n,n-1}} & iB_n & A_n^{(-)}e^{i\tilde{\Phi}_{n,n-1}} \\ iB_n e^{-i\Phi_{n,n-1}} & A_n^{(+)} + A_n^{(-)} & iB_n e^{i\tilde{\Phi}_{n,n-1}} \\ A_n^{(-)}e^{-i\Phi_{n,n-1}} & iB_n & A_n^{(+)}e^{i\tilde{\Phi}_{n,n-1}} \end{bmatrix}, \qquad (8)$$

where $\Phi_{n,n-1} = (\omega_{v+1} - \omega_v)(\tau_n - \tau_{n-1})$ and $\tilde{\Phi}_{n,n-1} = (\omega_v - \omega_{v-1})(\tau_n - \tau_{n-1})$. The matrix elements are defined by $A_n^{(\pm)} = (\pm 1 + \cos\sqrt{2}a_n)/2$ and $B_n = \sin\sqrt{2}a_n/\sqrt{2}$, where $a_n = \langle v'|R_n|v\rangle \delta_{v',v\pm 1}$. Here, we only consider the $|\Delta v|=1$ Raman transitions because of the small absolute values of the $|\Delta v|\geq 2$ Raman transition elements in the present system.

To verify Eq. (8), we apply Eq. (8) to the selective population transfer considered in Sec. IIIA with $v=30$ and $\psi_0 = (0,1,0)^t$. By expressing the Raman π pulse as a pulse train composed of identical $N$ pulses with the regular pulse interval $\Phi_{n,n-1} = (\omega_{31} - \omega_{30})T_{31,30} = 2\pi$, we assume the $n$-independent value of $a_n = -\pi/2N$ in Eq. (8), where the negative value comes from the present definition of the phases of the vibrational eigenstates and the factor 1/2 is also due to our definition. We plot the time evolution of the population of the state $|v=31\rangle$ by setting $N=33$, which is shown by the gray solid triangles in Figs. 3(b). This simplified treatment reasonably reproduces the time-dependent population induced by the optimal NR pulse and justifies the qualitative description of the vibrational dynamics by using Eq. (8) as a minimal model. For reference, we also calculate the time evolution of the population of the state $|v=31\rangle$ by assuming the simplified pulse train with $N=25$ in Fig. 4(b). As expected, such a simplified pulse train no longer reproduces the time-dependent behavior of the $|v=31\rangle$-state population, which again indicates the non-trivial control mechanisms when $t_f = 25 T_{31,30}$.

To qualitatively discuss the non-trivial control mechanisms in Fig. 4, it is convenient to simplify the expression up to the lowest-order terms with respect to the matrix elements $\{a_n\}$ although the lowest-order approximation cannot directly reproduce the entire population transfer. Under the approximations, it is straightforward to derive the probability amplitude of the $|v\pm 1\rangle$ state immediately after the $n$th pulse

$$C_{v+1}(\tau_n) = \sum_{j=1}^{n} a_j \left( \prod_{k=j+1}^{n} e^{-i\delta_k} \right) \text{ and } C_{v-1}(\tau_n) = \sum_{j=1}^{n} a_j \left( \prod_{k=j+1}^{n} e^{i\tilde{\delta}_k} \right) \quad (9)$$

where $\Phi_{k,k-1} \equiv 2\pi + \delta_k$ and $\tilde{\Phi}_{k,k-1} \equiv 2\pi + \tilde{\delta}_k$. Here, the phases $\delta_1$ and $\tilde{\delta}_1$ do not appear in Eq. (9) because of the initial state $\psi_0 = (0,1,0)^t$. To focus on the roles of the two types of pulse intervals in Fig. 4(a), we further assume the $n$-independent value of $a_n = a$ ($n=1,2,\cdots,N$). We then have the expression of the population of the $|v+1\rangle$ state as

$$P_{v+1}(\tau_n) = |a|^2 \{ n + 2(\cos\delta_2 + \cdots + \cos\delta_n) + 2[\cos(\delta_2 + \delta_3) + \cdots + \cos(\delta_{n-1} + \delta_n)] + \cdots + 2\cos(\delta_2 + \cdots + \delta_n) \} \quad (10)$$

If we replace $\{\delta_j\}$ with $\{\tilde{\delta}_j\}$ in Eq. (10), we will have the expression of the population of the $|v-1\rangle$ state, $P_{v-1}(\tau_n)$.

We first consider a simple case in which a pulse appears at every $T_{31,30}$, that is, $\{\delta_j = 0\}$, corresponding to the optimal pulse train in Fig. 3(a) ($t_f = 33 T_{31,30}$). The substitution of $v = 30$ into Eq. (10) yields the target population $P_{31}(\tau_n) = |a|^2 n^2$, which is proportional to the square of the number of pulses. The purely constructive quantum interference plays an active role in selectively achieving the population transfer to the target state $|v=31\rangle$. For the unwanted population $P_{29}(\tau_n)$, on the other hand, the phase shifts $\{\tilde{\delta}_j = 0.19\}$ are constantly added to the population of the state $|v=29\rangle$ at every $T_{31,30}$. The accumulated phase shifts weaken the constructive interference, turning it into the destructive interference, and finally leading to $P_{29}(\tau_{N=33}) \simeq 0$ when $t_f = 33 T_{31,30}$.

In the case of $t_f = 25 T_{31,30}$ (Fig. 4), it may be convenient to start our discussion by examining the unwanted population $P_{29}(\tau_{N=25}) \simeq 0$. To realize $P_{29}(\tau_{N=25}) \simeq 0$, we need to introduce larger phase shifts into $P_{29}(\tau_n)$ than those in the case of $t_f = 33 T_{31,30}$, to more quickly realize the destructive interference. Once the destructive phase shifts are adequately accumulated in the first half of the control period, the optimal pulse train tends to keep the destructive phase shifts in the second half by setting small values of $\{|\tilde{\delta}_j|\}$ to finally lead to $P_{29}(\tau_{N=25}) \simeq 0$. The introduction of such large phase shifts, however, inevitably results in the introduction of some phase shifts even into the target population $P_{31}(\tau_n)$, which makes the constructive interference effects in $P_{31}(\tau_n)$ insufficient. We thus need to adopt more intense pulses to make up for the insufficient constructive interferences to achieve $P_{31}(\tau_{N=25}) \simeq 1$. This could explain the increase in the pulse fluence to achieve the population transfer when $t_f < 33 T_{31,30}$ in Fig. 2.

**C. Wave packet shaping**

We consider a more general objective than the selective population transfer, that is, the wave-packet shaping control, in which we need to adjust both the population distribution and the relative phases, simultaneously. The initially excited state is assumed to be the vibrational eigenstate $|\psi(t_0 = 0)\rangle = |v = 30\rangle$. For the sake of systematic analyses, we consider the target wave packet

$$|\chi(\theta)\rangle = \frac{1}{\sqrt{3}}\left(e^{i\theta}|29\rangle + |30\rangle + |31\rangle\right), \tag{11}$$

which is expressed as a function of the relative phase $\theta$. As our purpose is to create the target wave packet in Eq. (11) at a specified final time $t_f$, the target operators are set to $X = |\chi(\theta)\rangle\langle\chi(\theta)|$ and $Y(t) = 0$ in Eq. (2). We measure the time in units of the averaged vibrational period $\bar{T} = (T_{31,30} + T_{30,29})/2$ for convenience. We assume a slightly longer final time $t_f = 60\bar{T}$ to avoid the restriction due to the lack of a control period.

We numerically design the (dimensionless) optimal pulse envelopes $f(t)$ while systematically changing the values of $\theta = 0, \pi/2, \pi$, and $3\pi/2$, as shown in Fig. 5. The optimal pulse envelopes have the same temporal structure but are temporally shifted according to the value of $\theta$. The $\theta$-dependent temporal shifts mean that the relative phases in the target wave packet are mainly adjusted by the free time propagation of the wave packet after the NR pulse control, which can explain the presence of the two optimal solutions in Fig. 5(a). The temporal shift between the two optimal solutions in Fig. 5(a) corresponds to the so-called revival time of the vibrational wave packet $\sim 33\bar{T}$ due to the anharmonicity of the potential. Note that in many case studies that adopt resonant laser pulses as control knobs, we also see such cooperation between the control laser pulse and the wave packet motion to adjust the relative phases, thereby achieving the control objectives. One example is the vibrational wave packet prepared by the negatively chirped pump pulse in the electronically excited state, which becomes a spatially localized wave packet after freely propagating in the anharmonic potential [48, 49].

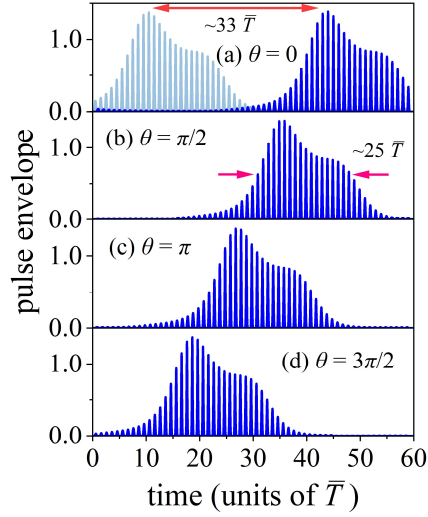

Figure 5
Optimal pulse envelopes that create the target wave packets as a function of time defined by Eq. (11) with (a) $\theta=0$, (b) $\theta=\pi/2$, (c) $\theta=\pi$, and (d) $\theta=3\pi/2$. Time is measured in units of $\bar{T}=(T_{31,30}+T_{30,29})/2$. In the case of (a) $\theta=0$, another optimal solution exists as shown by the light blue line. The interval between the two optimal solutions $\sim 33\bar{T}$ is indicated by a red double arrow. In (b), the overall temporal width of the optimal pulse train $\sim 25\bar{T}$ is indicated.

We next consider the control mechanisms regarding the population distribution. As the population distribution of the present target wave packet in Eq. (11) is independent of $\theta$, we focus on the $\theta$-independent features of the optimal pulse trains such as the overall temporal width $\sim 25\bar{T}$ and the appearance of each pulse at the regular interval $\sim \bar{T}$. To examine these two features, we adopt a model pulse train that is composed of 25 identical 80-fs Gaussian pulses with a regular pulse interval. Then, we calculate the populations of the vibrational eigenstates as a function of the pulse interval and the relative pulse fluence with respect to the fluence of the

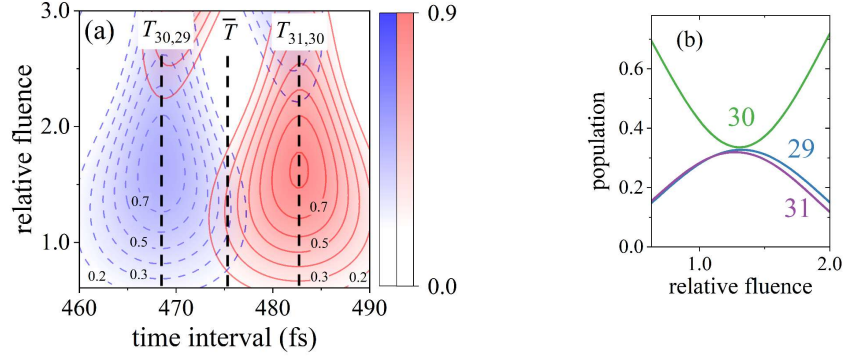

Figure 6
(a) Contour map of the populations of states $|v=31\rangle$ and $|v=29\rangle$ as a function of the pulse interval and the relative fluence with respect to the total fluence of the optimal pulse in Fig. 5(a). Here, the populations are calculated by using the model pulse train composed of equally distributed 25 identical 80-fs Gaussian pulses when the initially excited state is set to $|v=30\rangle$. (b) Cut of the contour map along the fixed pulse interval $\bar{T}=(T_{31,30}+T_{30,29})/2$, in which the populations of the three states are shown as a function of the relative fluence.

optimal pulse train in Fig. 4(a). Figure 6(a) shows a contour map of the populations of states $|v=31\rangle$ and $|v=29\rangle$. For reference, when the pulse intervals are set to $T_{31,30}$ and $T_{30,29}$, 90% of the populations at most are distributed in the states $|v=31\rangle$ and $|v=29\rangle$, respectively, reflecting the lack of frequency resolution due to the short overall temporal width $\sim 25\bar{T}$. If the pulse interval is set to $\bar{T}$, the three vibrational eigenstates in the target wave packet [Eq. (11)] are almost equally distributed, which is clearly illustrated by the cut of the contour map along $\bar{T}$ in Fig. 6(b). The results in Fig. 6 strongly suggest that the optimal pulses actively utilize the lack of frequency resolution to excite multiple vibrational eigenstates simultaneously to realize the target population distribution.

### D. Suppressing wave packet deformation

We quantify the degree of wave packet deformation by using the absolute square of the overlap integral $|\langle\psi_0|\psi(t)\rangle|^2$, where $|\psi_0\rangle=\sum_v |C_v| e^{-i\theta_v}|v\rangle$ is the initially excited state (Appendix A). In the absence of the control pulse, the temporal behavior of $|\langle\psi_0|\psi(t)\rangle|^2$ is solely dependent on the magnitude of the coefficients $\{|C_v|\}$ and independent of the initial phases

$\{\theta_v\}$. Roughly speaking, it shows a damped oscillation due to the relative-phase mismatches caused by the anharmonic potential. Our control objective in this subsection is to suppress the wave packet deformation such that the wave packet simply repeats the regular oscillation as if it were moving in a harmonic potential. We will examine how effectively we can suppress the wave packet deformation by focusing on the effects of the initial relative phases. The suppression of the wave packet deformation is nonintuitive in the sense that we have to adjust the relative phases of the eigenstates in the wave packet by using a moderately intense NR pulse while avoiding the population redistribution. We also note that the intensities of the NR pulses considered here are not sufficiently strong to directly manipulate the shape of the potential curve.

From the perspective of experimental feasibility, we adopt chirped pump pulses [50] to prepare various initially excited states $|\psi_0\rangle = \sum_v |C_v| e^{-i\theta_v} |v\rangle$. This is because the chirped pump pulses solely change the phases $\{\theta_v\}$ without changing $\{|C_v|\}$, which is suitable for the above-mentioned purpose. The Fourier transform of the (linearly) chirped pulse $\tilde{E}_{\text{pump}}(\omega)$ is defined by

$$\tilde{E}_{\text{pump}}(\omega) = \tilde{E}_{\text{pump}}^{\text{G}}(\omega) \exp\left[i\frac{\varphi_2}{2}(\omega - \omega_0)^2\right], \tag{12}$$

where $\omega$ and $\varphi_2$ are the angular frequency and the linear chirp rate, respectively. Here, $\tilde{E}_{\text{pump}}^{\text{G}}(\omega)$ is the Fourier transform of the Gaussian TL pump pulse with the central frequency $\omega_0$ and the temporal width (FWHM) of the intensity $2\sigma\sqrt{\ln 2}$. In the following numerical examples, we assume an 80-fs FWHM Gaussian TL pump pulse and adopt several values of $\varphi_2$ to prepare the initially excited states $|\psi_0\rangle$ during the period $[t_i = -300 \text{ fs}, t_0 = 300 \text{ fs}]$. When the central wavelength is set to 535 nm, $|\psi_0\rangle$ is mainly composed of the vibrational eigenstates $|v=29\rangle$, $|v=30\rangle$, and $|v=31\rangle$ with minor contributions from $|v=28\rangle$ and $|v=32\rangle$. In this subsection, the initially excited wave packets $|\psi_0\rangle$ created by TL, the negatively chirped pump pulses, and the positively chirped pump pulses will be referred to as the TL, NC, and PC wave packets, respectively. Some numerical examples of $|\psi_0\rangle$ are shown in Appendix A.

The target operators in Eq. (2) are set to $X = |\psi_0\rangle\langle\psi_0|$ and $Y(t) = |\psi_0\rangle y(t)\langle\psi_0|$,

where the positive envelope function $y(t)$ specifies the timings for evaluating $|\langle\psi_0|\psi(t)\rangle|^2$ during the control period $[t_0 = 300\,\text{fs}, t_\text{f}]$. Here, we assume that $y(t)$ is expressed as the sum of the Gaussian functions

$$y(t) = \frac{1}{\Delta t}\sum_{n=1}^{N} \exp\left[-\frac{(t-\tau_n)^2}{d^2}\right], \qquad (13)$$

where the set of $\{\tau_n\}$ specifies the peak positions. The peak height and the temporal width of each Gaussian function in Eq. (13) are set to the inverse of the numerical temporal grid $1/\Delta t = 1/0.1\,\text{fs}$ and $d = 8.5\times 10^{-3}\,\text{fs}$, respectively. The small value of $d$ is used to evaluate $|\langle\psi_0|\psi(t)\rangle|^2$ only around a specified set of $\{\tau_n\}$. In practice, the values of $\{\tau_n\}$ are updated in each iteration step of the solution algorithm, that is, if $|\langle\psi_0|\psi^{(k)}(t)\rangle|^2$ has maximal values at $\{\tau_n^{(k)}\}$ with $|\psi^{(k)}(t)\rangle$ being the wave packet at the $k$th iteration step, we substitute $\{\tau_n^{(k)}\}$ into Eq. (13) to define new peak positions used in the next ($k$+1)th iteration step. From the maximal values of $|\langle\psi_0|\psi^{(k+1)}(t)\rangle|^2$, we obtain the revised peak positions $\{\tau_n^{(k+1)}\}$ used in the ($k$+2)th iteration step, and so on. As shown in Appendix B, we numerically solve the inhomogeneous differential equation associated with the Lagrange multiplier instead of introducing the approximation method [24].

We numerically optimize the NR pulse envelopes for several values of $|\varphi_2|\leq 2.0$ measured in units of $\sigma^2 = \left(80\,\text{fs}/2\sqrt{\ln 2}\right)^2$. The final time is set to $t_\text{f} = 17\bar{T}$, which corresponds to $N=16$ in Eq. (13), where the time is measured in units of the average vibrational period $\bar{T} = (T_{31,30} + T_{30,29})/2 = 475\,\text{fs}$. The results are summarized in Fig. 7 by using the degree of "deformation suppression", which is defined by the averaged value

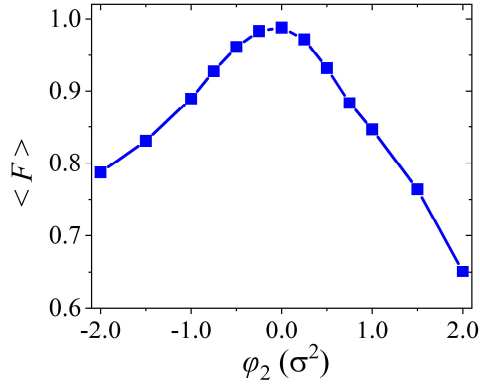

Figure 7

Degree of deformation suppression $<F>$ defined by Eq. (14) as a function of the linear chirp rate $\varphi_2$ which is measured in units of the Gaussian pump pulse width $\sigma^2 = \left(80\,\text{fs}/2\sqrt{\ln 2}\right)^2$. Here, the control time is set to $t_\text{f} = 17\bar{T}$ with $\bar{T} = (T_{31,30} + T_{30,29})/2 = 475$ fs.

$$<F> = \frac{1}{N+1} \sum_{n=1}^{N+1} \left|\langle\psi_0|\psi(\tau_n)\rangle\right|^2, \qquad (14)$$

with $\tau_{n=N+1} \equiv t_\text{f}$. According to the definition in Eq. (14), perfect deformation suppression leads to $<F> = 1$. On the other hand, $<F> = 0.56$ in the absence of the control pulse. We see from Fig. 7 that the values of $<F>$ are greater than 0.95 when $-0.5 < \varphi_2 < 0.25$. We also see an asymmetric structure, that is, there is a lack of symmetry with respect to $\varphi_2$. When the value of $|\varphi_2|$ exceeds the above range, the deformation suppression control becomes less effective although the optimal NR pulses always suppress the deformation of the wave packets to some degree.

To consider the $\varphi_2$-dependent control mechanisms, we examine the results focusing on some typical values of $\varphi_2$. Figure 8 shows the results when $\varphi_2 = 0$ fs$^2$ (TL pump pulse). The degree of deformation suppression [Eq. (14)] has a value of $<F> = 0.988$, which is close to the ideal value. The optimal NR pulse in Fig. 8(a) is a pulse train composed of almost identical pulses that appear precisely at every $\bar{T}$ within $\pm 1$ fs (see the purple solid squares). The optimal NR pulse makes the TL wave packet oscillate regularly [Fig. 8(b)] as if the wave packet is in a harmonic potential. In addition, the population transfer is minimized to keep the wave packet in the original shape even under the influence of mildly intense laser pulses [Fig. 8(c)].

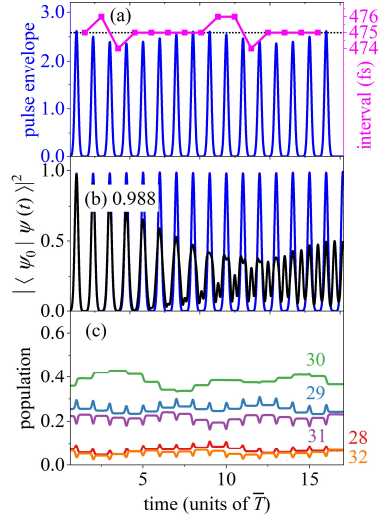

Figure 8

Deformation suppression control when $t_f = 17\bar{T}$ in the case of $\varphi_2 = 0 \text{ fs}^2$. (a) Optimal pulse envelope in which the intervals between the adjacent pulses are shown by purple solid squares. The black dashed line shows $\bar{T} = 475 \text{ fs}$. (b) Degree of deformation suppression is $<F> = 0.988$. Absolute square of the overlap integral, $|\langle \psi_0 | \psi(t) \rangle|^2$ in the presence (absence) of the control pulse is shown by the blue (black) line. (c) Populations of the five major states as a function of time (units of $\bar{T}$).

It is also useful to see the results when the deformation suppression control is not so effective owing to the increase in the value of $|\varphi_2|$. As an example, we show the results when the chirp rates are set to $\varphi_2 = -1.0 \, \sigma^2$ (Fig. 9) and $\varphi_2 = +1.0 \, \sigma^2$ (Fig. 10), which lead to the optimal values of $<F> = 0.889$ and $<F> = 0.846$, respectively. The reduction of the values of $<F>$ may be qualitatively attributed to the spatially delocalized structures of $|\psi_0\rangle$. That is, the wave packets are controlled by the nuclear-coordinate-dependent polarizability interaction, which may not be favorable for some parts of the wave packets to suppress the deformation. Scrutinizing the results in Fig. 9, the first two pulses [Fig. 9(a)] are so weak that they essentially let the NC wave packet freely move, suggesting the limited capability of the polarizability interaction to keep the original shape of the NC wave packet. During the (almost) free propagation, the NC wave packet is modified such that it is spatially more localized than $|\psi_0\rangle$ and approaches the shape of the TL wave packet. The localized feature of the modified wave

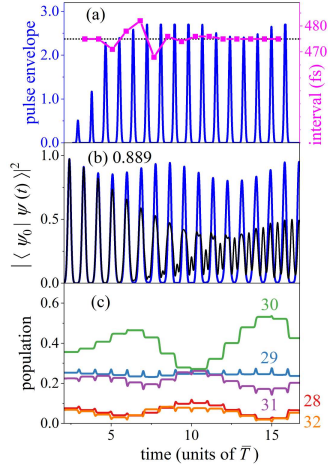

Figure 9

Deformation suppression control when $t_f = 17\bar{T}$ in the case of $\varphi_2 = -1.0\sigma^2$, where $\sigma^2 = \left(80\,\text{fs}/2\sqrt{\ln 2}\right)^2$. (a) Optimal pulse envelope in which the intervals between the adjacent pulses are shown by purple solid squares. The black dashed line shows $\bar{T} = 475\,\text{fs}$. (b) Degree of deformation suppression is $<F> = 0.889$. Absolute square of the overlap integral, $\left|\langle\psi_0|\psi(t)\rangle\right|^2$ in the presence (absence) of the control pulse is shown by the blue (black) line. (c) Populations of the five major states as a function of time (units of $\bar{T}$).

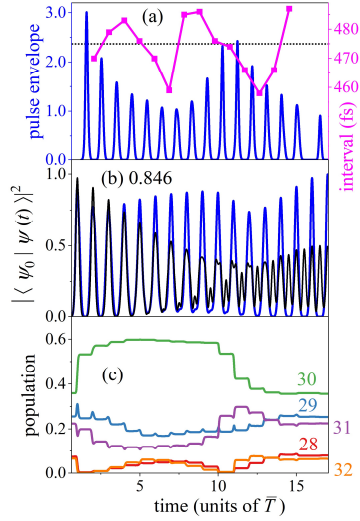

Figure 10

Deformation suppression control when $t_f = 17\bar{T}$ in the case of $\varphi_2 = +1.0\sigma^2$, where $\sigma^2 = \left(80\,\text{fs}/2\sqrt{\ln 2}\right)^2$. (a) Optimal pulse envelope in which the intervals between the adjacent pulses are shown by purple solid squares. The black dashed line shows $\bar{T} = 475\,\text{fs}$. (b) Degree of deformation suppression is $<F> = 0.846$. Absolute square of the overlap integral, $\left|\langle\psi_0|\psi(t)\rangle\right|^2$ in the presence (absence) of the control pulse is shown by the blue (black) line. (c) Populations of the five major states as a function of time (units of $\bar{T}$).

packet would make it easier to suppress the deformation while leading to the loss of the original shape. These contradictory factors would lead to the optimal solution in Fig. 9 as a compromise.

The initial relative phases of the PC wave packet are consistent with the anharmonic potential to accelerate the deformation of the PC wave packet. Contrary to the NC wave packet, the PC wave packet in Fig. 10 is on the way to the so-called collapse. Therefore, the control pulse should stop the spatial delocalization of the PC wave packet as soon as possible, which explains the first intense pulse in Fig. 10(a). As shown in Fig. 10(b), this feature of the PC wave packet makes the deformation suppression even more difficult than that in Fig. 9(b), leading to the highly structured optimal NR pulse envelope with irregular pulse intervals [Fig. 10(a)]. The very last pulse in particular is π-phase shifted and appears when the wave packet is localized around the outer turning point. In fact, the optimal NR pulse in Fig. 10(a) induces a large temporal change in the populations in Fig. 10(c). In this regard, the control in Fig. 10 would partly include the wave packet reshaping processes in addition to the deformation suppression.

So far, we have restricted ourselves to the cases where the initially excited states are prepared by the chirped pump pulses. We now consider the initial-phase dependence of the degree of deformation suppression $<F>$ by adopting the three-state model under the impulsive excitation approximation (Sec. IIIB). The initial state is generally expressed as

$$\psi_0 = \begin{bmatrix} |C_{v+1}^0| e^{-i\theta} \\ |C_v^0| \\ |C_{v-1}^0| e^{i\tilde{\theta}} \end{bmatrix} \quad (15)$$

where $\theta = \theta_{v+1} - \theta_v$ and $\tilde{\theta} = \theta_v - \theta_{v-1}$ with $\theta_v$ being the initial phase associated with the vibrational eigenstate $|v\rangle$, etc. In Eq. (15), we remove the phase $\theta_v$ as a global phase. Although the condition for suppressing the deformation, for instance, immediately after the first pulse would be given by $|\psi_0^\dagger \psi(\tau_1)|^2 = 1$, we could not derive the approximate expression that is useful to examine the control mechanisms. We thus directly apply Eq. (8) to the model to calculate $<F>$ as a function of $\theta$ and $\tilde{\theta}$ in the range of $-\pi \leq \theta, \tilde{\theta} < \pi$.

Referring to the results in Fig. 8, we assume an NR pulse train composed of 17 identical pulses, the total pulse area of which is set to the Raman π pulse. The fixed pulse interval

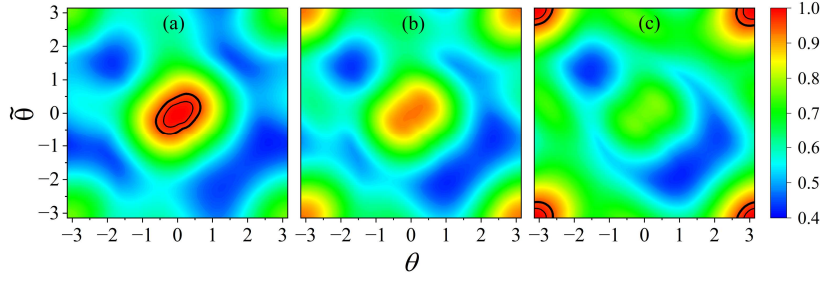

Figure 11

Contour plots of the degree of deformation suppression $<F>$ derived from the three-state model as a function of the initial relative phases $\theta$ and $\tilde{\theta}$ [see Eq. (15)]. The sets of the absolute values of the coefficients are given by (a) $(|C_{v+1}|\ |C_v|\ |C_{v-1}|) = (\sqrt{1/6}\ \sqrt{2/3}\ \sqrt{1/6})$, (b) $(1/2\ \sqrt{1/2}\ 1/2)$, and (c) $(\sqrt{1/3}\ \sqrt{1/3}\ \sqrt{1/3})$. In Figs. 11(a) and (c), the bold and thin black contour lines show the values of 0.98 and 0.95, respectively.

$\bar{T} = (T_{31,30} + T_{30,29})/2 = 475$ fs is assumed. On the basis of our definition of the phases of the vibrational eigenstates, we consider the three sets of coefficients $(|C_{v+1}|\ |C_v|\ |C_{v-1}|)$ with $v = 30$ in Fig. 11, that is, (a) $(\sqrt{1/6}\ \sqrt{2/3}\ \sqrt{1/6})$ that leads to the spatially localized wave packet around the inner turning point when $\theta = \tilde{\theta} = 0$ and (c) $(\sqrt{1/3}\ \sqrt{1/3}\ \sqrt{1/3})$ that leads to that around the outer turning point when $|\theta| = |\tilde{\theta}| = \pi$. As an example of the intermediate case between (a) and (c), we also consider (b) $(1/2\ \sqrt{1/2}\ 1/2)$ that does not lead to the spatially localized wave packet. Although the asymmetric distribution of $<F>$ is apparent in Fig. 7, we see almost the symmetric distributions of $<F>$ with respect to $\theta$ and $\tilde{\theta}$ in Fig. 11, which could be attributed to the simplified model analyses. In Fig. 11(a) [Fig. 11(c)], the wave packet deformation is highly suppressed in the region around $\theta = \tilde{\theta} = 0$ ($|\theta| = |\tilde{\theta}| = \pi$), which corresponds to the pulse irradiation timing when the wave packet is localized around the inner (outer) turning point. Contrary to the results in Figs. 11(a) and (c), the largest value of $<F>$ is 0.92 in the intermediate case [Fig. 11(b)]. Considering all the results in this subsection, we conclude that the NR pulse can suppress the deformation of the spatially localized wave packet if the mildly intense NR pulse is regularly applied to the wave packet at the time when it is spatially localized. For spatially delocalized wave packets, on the other hand, it would be difficult to suppress the deformation although shaped NR pulses could improve the degree of deformation suppression to some extent.

**IV. Summary**

On the basis of optimal control theory, we have discussed how accurately we can control the probability amplitudes of the vibrational eigenstates by using the cycle-averaged polarizability interactions induced by mildly intense NR pulses, which lack the control knobs originating from the optical frequencies and their relative phases. For this purpose, we have considered three kinds of control objectives: (i) the selective population transfer that solely requires the control of the absolute value of the probability amplitude, (ii) the wave packet shaping that requires both the population control and the relative-phase control, and (iii) the suppression of wave packet deformation that solely requires the control of the relative phases without changing the population distribution. From the perspective of experimental feasibility, we have considered the vibrational dynamics in the electronically excited $B$ state of $I_2$ as a case study because we can precisely prepare the specified initial states by (shaped) pump pulses. We have numerically designed the optimal NR pulses associated with the three kinds of control objectives and examined the control mechanisms with the aid of the three-state model under the impulsive excitation approximation. The results are summarized as follows.

(i) When the control period is sufficiently long to frequency-resolve the target Raman transition, the optimal solution is a Raman $\pi$ pulse composed of almost identical pulses with regular intervals. Interestingly, even if the control periods are not sufficiently long to resolve the Raman transition frequency, the optimally shaped NR pulses achieve almost complete selective population transfer. In fact, we have numerically shown that the control periods can be shortened to less than one-third of that expected from the frequency resolution. The shaped optimal pulses adjust the pulse intervals to actively introduce constructive and destructive quantum interferences to the target state and the unwanted state, respectively, although the pulse fluence monotonically increases with the decrease in the control period. An extremely high pulse fluence may impose restrictions on the experimental feasibility.

(ii) The optimal pulse is the pulse train, the overall temporal width of which is adjusted

to induce multiple population transitions due to the lack of frequency resolution, whereby the target population distribution is achieved. On the other hand, the relative phases are largely adjusted through the free propagation in the anharmonic potential, which is similar to the way of shaping wave packets, etc. with resonant laser pulses[48, 49].

(iii) In the present study of the suppression of wave packet deformation, we have focused on the effects of the relative phases { $\theta_v$ } in the initial wave packet $|\psi_0\rangle = \sum_v |C_v| e^{-i\theta_v} |v\rangle$, which is prepared by the chirped pump pulses so that a fixed value of {$|C_v|$} is assumed. From the simulation and the model analyses, we have shown that the pulse trains that are composed of almost identical pulses that appear regularly in every vibrational period can suppress the deformation in a long control period with a high probability, provided that the wave packets are spatially localized. As the degree of the spatial delocalization of the initially excited wave packets is increased, the effectiveness of the deformation suppression control is gradually reduced to a certain extent although the shaped NR pulses could improve the deformation suppression to some degree.

From the successful control systematically demonstrated by focusing on the vibrational dynamics, we conclude that the shaped NR laser pulses can be as effective as the shaped resonant laser pulses to nonadiabatically manipulate the probability amplitudes of multi-level quantum systems. This would increase the degree of freedom for choosing laser pulses to manipulate quantum dynamics for achieving the specified control objectives with high accuracy.

**Acknowledgments**

One of the authors (YO) acknowledges support in the form of Grants-in-Aid for Scientific Research (C) (20K05414 and 23K04659) and partly from the Joint Usage/Research Program on Zero-Emission Energy Research, Institute of Advanced Energy, Kyoto University (ZE2023B-08). This work was supported by MEXT Quantum Leap Flagship Program (MEXT Q-LEAP) JPMXS0118069021 and JSPS Grant-in-Aid for Specially Promoted Research Grant No. 16H06289.

**Appendix A: Initially excited state $|\psi_0\rangle$ in the $B$ state and numerical details**

We assume a two-electronic-state model that consists of the $X$ (ground) and $B$ states of $I_2$, the vibrational wave packets of which are given by $|\psi_X(t)\rangle$ and $|\psi_B(t)\rangle$, respectively. They obey the Schrödinger equation

$$i\hbar \frac{\partial}{\partial t} \begin{bmatrix} |\psi_B(t)\rangle \\ |\psi_X(t)\rangle \end{bmatrix} = \begin{bmatrix} H_B(t) & -\mu_{BX}(r)E_{\text{pump}}(t) \\ -\mu_{XB}(r)E_{\text{pump}}(t) & H_X(t) \end{bmatrix} \begin{bmatrix} |\psi_B(t)\rangle \\ |\psi_X(t)\rangle \end{bmatrix}, \quad \text{(A1)}$$

where $H_X(t)$ and $H_B(t)$ are the vibrational Hamiltonians that include the polarizability interactions. The electronic transition is induced by the pump pulse $E_{\text{pump}}(t)$ with the electric dipole moment function $\mu_{BX}(r) = [\mu_{XB}(r)]^\dagger$ with $r$ being the internuclear distance. Assuming that the molecule is initially in the lowest vibrational state in the $X$ state $|\psi_X(t_i)\rangle = |0_X\rangle$ (see Fig. 1) and that the pump pulse is sufficiently weak to be treated by the first-order perturbation, we have

$$|\psi_B(t_0)\rangle \simeq \frac{i}{\hbar} \int_{-\infty}^{\infty} ds\, e^{-i(t_0-s)H_B^0/\hbar} \mu_{BX}(r) E_{\text{pump}}(s)\, e^{-i(s-t_i)H_X^0/\hbar} |0_X\rangle \equiv e^{-iH_B^0 t_0/\hbar} |\psi_B^0\rangle \quad \text{(A2)}$$

at $t = t_0$ (after the pump pulse) when the control by the NR pulse starts. Here, $H_X^0$ and $H_B^0$ describe the field-free Hamiltonians. We will call $|\psi_0\rangle = |\psi_B^0\rangle \big/ \sqrt{\langle \psi_B^0 | \psi_B^0 \rangle}$ the initially excited state in the present study.

We assume the rotating wave approximation (RWA) and numerically integrate Eq. (A2) with the temporal grid size $\Delta t = 0.1$ fs to obtain the (normalized) initial state $|\psi(t_0)\rangle = e^{-iH_B^0 t_0/\hbar} |\psi_0\rangle$ in Eq. (1). We adopt the same molecular parameters as those used in our previous study, namely, the potential energies and the transition dipole moment function are taken from Refs. [51-53]. We assume the polarizability interaction function in the $B$ state in Ref. [46]. The time evolution of the wave packets is calculated by using the split-operator method combined with fast Fourier transform (FFT) [3, 54, 55], in which the spatial range [2.1 Å, 6.0 Å]

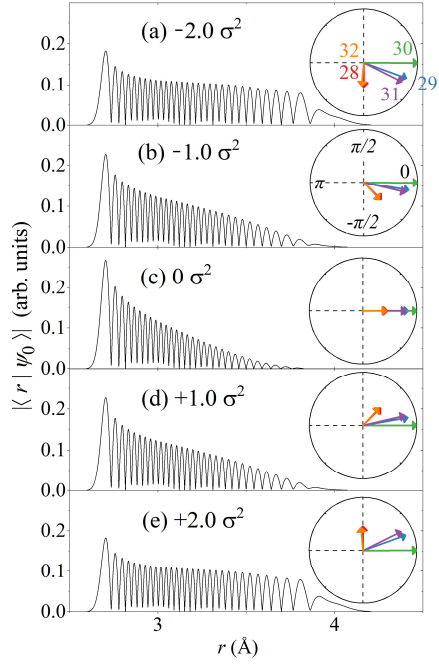

Figure 12

Examples of the initially excited states when the linear chirp rates are set to (a) $\varphi_2 = -2.0\sigma^2$, (b) $\varphi_2 = -1.0\sigma^2$, (c) $\varphi_2 = 0$ (TL pump pulse), (d) $\varphi_2 = +1.0\sigma^2$, and (e) $\varphi_2 = +2.0\sigma^2$ $\varphi_2 = -2.0\sigma^2$, where $\sigma^2 = \left(80\,\text{fs}/2\sqrt{\ln 2}\right)^2$. Here, $\varphi_2 < 0$ and $\varphi_2 > 0$ correspond to negatively and positively chirped pump pulses, respectively. The arrows in each inset circle schematically illustrate the probability amplitudes of the vibrational eigenstates in the initially excited state in the complex plane. The radius of the circle is set to the magnitude of the probability amplitude of the state $|v = 30\rangle$.

is equally divided into 512 grid points. Some numerical examples of the initially excited states are shown in Fig. 12.

**Appendix B: Backward time evolution of the Lagrange multiplier in Eq. (4)**

When we numerically calculate the time evolution of the vibrational wave packet, we often adopt a procedure that combines the split-operator method and FFT because the latter considerably reduces the computational cost [3, 54, 55]. When calculating the time evolution of the Lagrange multiplier in Eq. (4), we may have to modify the procedure to deal with the inhomogeneous term. Here, we examine the modification. Note that Eq. (4) describes the time backward evolution because the final condition is specified. We may introduce $s = t_f - t$ to rewrite Eq. (4) in the form of the time forward propagation:

$$\frac{\partial}{\partial s}|\bar{\xi}(s)\rangle = \frac{i}{\hbar}\bar{H}_B^\dagger(s)|\bar{\xi}(s)\rangle + \bar{Y}(s)|\bar{\psi}(s)\rangle, \tag{B1}$$

where $|\bar{\xi}(s)\rangle = |\xi(t_f - s)\rangle$, $\bar{H}_B^\dagger(s) = H_B^\dagger(t_f - s)$, $\bar{Y}(s) = Y(t_f - s)$, and $|\bar{\psi}(s)\rangle = |\psi(t_f - s)\rangle$. The "initial" condition is given by $|\bar{\xi}(0)\rangle = |\xi(t_f)\rangle$.

We try to find the solution having the form of $|\bar{\xi}(s)\rangle = \bar{U}(s,0)|\bar{g}(s)\rangle$ where the time evolution operator is defined by

$$\bar{U}(s,0) = \hat{T}\exp\left[\frac{i}{\hbar}\int_0^s ds'\, \bar{H}_B^\dagger(s')\right] \tag{B2}$$

with the time-ordering operator $\hat{T}$. The equation of motion for $|\bar{g}(s)\rangle$ is expressed as

$$\frac{\partial}{\partial s}|\bar{g}(s)\rangle = \bar{U}^{-1}(s,0)\bar{Y}(s)|\bar{\psi}(s)\rangle, \tag{B3}$$

where the initial condition is $|\bar{g}(0)\rangle = |\bar{\xi}(0)\rangle = |\xi(t_f)\rangle$. If we integrate both sides of Eq. (B3) over $[0, \Delta t]$, we have

$$|\bar{g}(\Delta t)\rangle - |\bar{g}(0)\rangle = \int_0^{\Delta t} ds\, \bar{U}^{-1}(s,0)\bar{Y}(s)|\bar{\psi}(s)\rangle \simeq \frac{\Delta t}{2}[\bar{U}^{-1}(\Delta t,0)\bar{Y}(\Delta t)|\bar{\psi}(\Delta t)\rangle + \bar{Y}(0)|\bar{\psi}(0)\rangle], \tag{B4}$$

where $\Delta t$ is the temporal grid size. Here, we adopt the trapezoid formula to approximately calculate the integral. Substituting Eq. (B4) into $|\bar{\xi}(s)\rangle = \bar{U}(s,0)|\bar{g}(s)\rangle$, we have

$$\left|\bar{\xi}(\Delta t)\right\rangle \simeq \bar{U}(\Delta t,0)\left[\left|\bar{\xi}(0)\right\rangle + \frac{\Delta t}{2}\bar{Y}(0)\left|\bar{\psi}(0)\right\rangle\right] + \frac{\Delta t}{2}\bar{Y}(\Delta t)\left|\bar{\psi}(\Delta t)\right\rangle. \tag{B5}$$

Repeating the procedure, we obtain

$$\left|\bar{\xi}(s_n)\right\rangle \simeq \bar{U}(s_n, s_{n-1})\left[\left|\bar{\xi}(s_{n-1})\right\rangle + \frac{\Delta t}{2}\bar{Y}(s_{n-1})\left|\bar{\psi}(s_{n-1})\right\rangle\right] + \frac{\Delta t}{2}\bar{Y}(s_n)\left|\bar{\psi}(s_n)\right\rangle \tag{B6}$$

for the $n$th step, where $s_n = s_{n-1} + \Delta t$. The time evolution operator $\bar{U}(s_n, s_{n-1})$ can be numerically calculated by using the standard split operator method in combination with FFT.